\documentclass[journal]{journal}

\ifCLASSINFOpdf
   \usepackage[pdftex]{graphicx}
\else
\fi

\usepackage[cmex10]{amsmath}
\usepackage{url}

\usepackage[numbers, sort]{natbib}
\usepackage{algorithm}
\usepackage{algpseudocode}
\usepackage{booktabs}
\usepackage{bm}
\usepackage[hidelinks]{hyperref}
\usepackage{subfigure}
\usepackage{amsfonts}

\newcommand{\Var}[1]{\mathrm{Var}\left(#1\right)}
\newcommand\norm[1]{\left\lVert#1\right\rVert} 
\DeclareMathOperator*{\argmin}{arg\,min}
\renewcommand*{\vec}[1]{\boldsymbol{#1}}

\begin{document}
%
\title{Analysing the Influence of Attack Configurations on the Reconstruction of Medical Images in Federated Learning}
%
%
%

\author{Mads~Emil~Dahlgaard,
        Morten~Wehlast~Jørgensen,
        Niels~Asp~Fuglsang,
        and~Hiba~Nassar
\thanks{M. E. Dahlgaard (\textit{s164206@student.dtu.dk}), M. W. Jørgensen, N. A. Fuglsang, and H. Nassar (\textit{hibna@dtu.dk}) are with Department of Applied Mathematics and Computer Science, Technical University of Denmark.}
\thanks{Code: \href{https://github.com/madsemildahlgaard/dlg-attack-configuration/}{\texttt{/madsemildahlgaard/dlg-attack-configuration/}}}
}


\maketitle
\thispagestyle{empty}

\begin{abstract}
The idea of federated learning is to train deep neural network models collaboratively and share them with multiple participants without exposing their private training data to each other. This is highly attractive in the medical domain due to patients' privacy records. However, a recently proposed method called Deep Leakage from Gradients enables attackers to reconstruct data from shared gradients. This study shows how easy it is to reconstruct images for different data initialization schemes and distance measures. We show how data and model architecture influence the optimal choice of initialization scheme and distance measure configurations when working with single images. We demonstrate that the choice of initialization scheme and distance measure can significantly increase convergence speed and quality. Furthermore, we find that the optimal attack configuration depends largely on the nature of the target image distribution and the complexity of the model architecture.
\end{abstract}


\section{Introduction}
In the modern era of deep learning, distributed learning has become a necessity for many applications when training on private datasets. Distributed or collaborative learning is a way for multiple parties to collaborate on a joint machine learning model without directly sharing private training data between parties. This could be the case for multiple hospitals with privacy-sensitive patient data working on a joint model.  As a result of the high importance of patient privacy, many medical researchers found distributed learning a good tool to benefit from machine learning techniques on one side and preserving privacy on the other side, see for example \cite{roy2019, sheller2018, rieke2020}. Instead of sharing training data, the model is distributed to each party, and only gradients from each party's training data are exchanged with either a parameter server (centralized approach) \cite{Iandola2016, Li2014} or neighbouring parties in the setup (decentralized approach) \cite{Patarasuk2009, Sergeev2018}. The model is then trained based on the shared gradients on a central server or locally and the process of sharing the joint model and individual gradients are repeated until training is satisfactory. This means that each party has its own private training data and never shares the actual data with other parties or a central server. 

Distributed learning makes the promise that each party can leverage the advantages of larger training sets while keeping their data private. It was previously thought that distributed learning is safe and that gradients do not reveal anything about the input data. However, recent studies show that it is indeed possible to expose the original training data based on gradients in certain circumstances. As an example, the algorithm Deep Leakage from Gradients (DLG) enables the reconstruction of training data from shared gradients in certain circumstances \cite{Zhu2020}. \citet{li2020} extends the algorithm to multi-label classifications for medical images. In this study, we show the effect of the initialization of the dummy variables and distance measures, that are used in the DLG algorithm, on the quality, speed, and stability of reconstructions. Recently, some researchers have explored defense strategies against attacks to prevent the leakage of important data \cite{li2020, ziller2021}. However, none of these defense strategies completely prevents leakage and in this work, we only discuss the attack possibilities in various scenarios.

The paper is organized as follows. We start in Section~\ref{sec:RW} with a brief account of related work. This is followed by Section~\ref{sec:theory} where  mathematical fundamentals of the algorithm are presented. Section~\ref{sec:experiments} demonstrate our results of the impact of dummy variables initialization and distance measure on the reconstruction quality, speed, and stability. The results are discussed in Section~\ref{sec:discussion}. Finally, the contribution of this study is concluded in Section~\ref{sec:conclusion}.


\section{Related Work}
\label{sec:RW}
\citet{BrendanMcMahan2017} introduced a decentralized approach for training on distributed devices called Federated Learning. This approach allows users to build a shared model based on private data without the need to centrally store the private data. They do so by leveraging the distributed mini-batch algorithm where each device only forwards the average gradient for their batch of data and not the actual data \cite{Dekel2012}. The average gradient of multiple devices is then used to perform a consensus update to the shared model. To demonstrate its practical application, Google used Federated Learning to predict user keyboard input \cite{BrendanMcMahan2017a}. Federated learning is also widely used in medical imaging, we name  \citet{feng, guo, wenqi2019, Sheller2020} among many.

The work on distributed and federated learning using gradient based algorithms builds on the assumption that gradients of a training epoch does not reveal information about the original data. However, \citet{Zhu2020} show that this is not the case by introducing the algorithm DLG. They demonstrate that one can reconstruct the initial training data, with pixel-wise accuracy on computer vision tasks, given only the model and gradients from a training batch. They do so by using supervised learning to update dummy input data with the goal of minimizing the distance between the dummy gradients and the true gradients. \citet{Wang2020} improve upon the results by introducing the SAPAG algorithm, which generalizes the concepts of DLG to multiple Deep Neural Network architectures. They use a Gaussian kernel for the distance measure between the dummy and the true gradients rather than Euclidean distance. \citet{Zhao2020} improve upon the results further by showing that the ground truth label for single label classification can be found directly from the gradients. They propose the improved Deep Leakage from Gradients method (iDLG) which improves the convergence rate of reconstructing single images from gradients by leveraging the knowledge of the true training data label.


\section{Theory} \label{sec:theory}
\subsection{Deep Leakage from Gradients (DLG)}
In federated learning, models are shared between all parties and trained locally and only the gradient is passed to the main server. DLG seeks to reconstruct an input $\vec{X}$ and a target $\vec{Y}$ given a twice differentiable model $F$, weights $\vec{W}$, and gradients $\nabla \vec{W}$.  The method can be described in the following steps:
\begin{enumerate}
  \item Initialize training data and their labels $(\vec{X}', \vec{Y}')$ randomly.
    \item Compute the dummy gradients $\nabla \vec{W}'$ of the model using  $(\vec{X}', \vec{Y}')$. 
    \item Solve as an optimization problem by finding the set $(\vec{X}^*, \vec{Y}^*)$ that minimizes the distance between the dummy gradients  $\nabla \vec{W}'$ and the shared one $\nabla \vec{W}$, i.e. 
\begin{equation}
    (\vec{X}^*, \vec{Y}^*) = \argmin_{(\vec{X}', \vec{Y}')}
    \mathcal{D}(\nabla \vec{W}', \nabla \vec{W}).
\end{equation}
\end{enumerate}
The updating procedure is presented in its general form in Algorithm \ref{DLGalgo} where $\pi_X$ and $\pi_Y$ specify initialization distributions, $\mathcal{D}$ is a distance measure, $N$ is the number of iterations, and $\eta$ denotes the learning rate. Our work builds on the general federated learning setting presented by \citet{Zhu2020}. Consequently, any twice differentiable model in a federated learning setting will be prone to the attacks discussed in this paper.


\begin{algorithm}
\caption{General algorithm for image reconstruction}\label{DLGalgo}
\begin{algorithmic}[1]
\Require: $F$ (twice differentiable), $\nabla \vec{W}$, $\vec{W}$
\State $\vec{X}'\gets \pi_X(\theta_X), \vec{Y}'\gets\pi_Y(\theta_Y)$
\For{$i \gets 1 \dots N$}
\State $\nabla \vec{W}' \gets \frac{\partial \mathcal{L}(F(\vec{X}', \vec{W}), \vec{Y}')}{\partial W}$
\State $L_G = \mathcal{D}(\nabla \vec{W}', \nabla \vec{W})$
\State $\vec{X}'\gets \vec{X}' - \eta \nabla_{\vec{X}'}L_G$
\State $\vec{Y}'\gets \vec{Y}' - \eta \nabla_{\vec{Y}'}L_G$
\EndFor
\end{algorithmic}
\end{algorithm}

\subsection{Initialization of dummy data}
In this work we consider two different ways of initializing the dummy image, $\vec{X}'$, and the dummy target, $\vec{Y}'$, to compare between. Following \citet{Qian2020}, a uniform distribution $\hat{\vec{X}}, \hat{\vec{Y}} \sim \mathcal{U}(0,1)$ is studied. The second distribution we study is a standard normal distribution $\hat{\vec{X}}, \hat{\vec{Y}} \sim \mathcal{N}(\vec{0}, \vec{I})$. Both distributions are scaled to the range [0;1] such that
\begin{equation}\label{eq:dist_scaling}
    \vec{X}' = \frac{\hat{\vec{X}} - \hat{\vec{X}}_{min}}{\hat{\vec{X}}_{max}-\hat{\vec{X}}_{min}},
\end{equation}
and similarly for $\vec{Y}'$. This is under the assumption that the ranges of the input image and target are [0;1]. This scaling can be seen as standardizing the range of the input dummy data that are generated from Gaussian random variables.
This suggestion is in an attempt to get the initial dummy inputs, $\vec{X}', \vec{Y}'$, closer to the true values $\vec{X}, \vec{Y}$. The scaled uniform initialization scheme is denoted by Unif and the scaled normal initialization scheme is denoted by Transformed Gaussian (TG) in the following sections. Consequently, in the following experiments we have that $\vec{X'}, \vec{Y'}\in[0,1]$.

\subsection{Distance measure for gradient matching}
\citet{Zhu2020} proposed using the Euclidean (Eucl) distance measure for gradient matching
\begin{align}
    \mathcal{D}(\nabla \vec{W}', \nabla \vec{W}) = \norm{\nabla \vec{W}' - \nabla \vec{W}}^2.
\end{align}
Trying to improve upon DLG, \citet{Wang2020} proposed using a weighted Gaussian measure for each layer given by
\begin{equation}
 \label{eq:sapag}
    \mathcal{D}\left(\nabla \boldsymbol{W}^{\prime}, \nabla \boldsymbol{W}\right)=Q\left(1-\exp \left[\frac{-\left\|\nabla \boldsymbol{W}^{\prime}-\nabla \boldsymbol{W}\right\|^{2}}{\lambda^{2}}\right]\right),
\end{equation}
where the factor $Q$ specifies the weight of a particular layer and $\lambda^2 = \Var{\nabla \vec{W}}$ is proposed as the optimal estimate of $\lambda^2$. The Gaussian kernel considers not only the element-wise relationship between gradient vectors, but also non-linear relationships.

In practice we found setting $\lambda^2=\Var{\nabla \vec{W}}$ to be unstable for single images (see appendix \ref{app:sigma_choice}). Therefore, we adopt a different approach with $\lambda^2$ given by $\lambda^2 = n\Var{\nabla\vec{W}}$, where $n$ is the number of parameters in the model layer. 
We denote this measure the Adaptive Gaussian measure (AG). The AG measure was found to perform better than the proposed Gaussian measure empirically on single images and is explored further in Section \ref{sec:sigma}.

\subsection{Similarity measures}
To evaluate the quality of the reconstructed image we measure how similar the reconstructed image is to the original image. We use two similarity measures: mean squared error (MSE) and structural similarity index measure (SSIM). 
MSE measures the mean of squared pixel-wise differences between the reconstructed image, $\vec{X}'$, and the true image, $\vec{X}$, as
\begin{equation}
    \text{MSE}(\vec{X}', \vec{X}) = \frac{1}{M} \sum_{i=1}^M (\vec{X}'_i - \vec{X}_i)^2,
\end{equation}
where $M$ is the total number of pixels. The mean over channels is calculated for multichannel images.

SSIM takes luminance, contrast, and structure into account, calculated as
\begin{equation}
    \text{\small SSIM}(\vec{X}', \vec{X}) = \frac{(2\mu_{\vec{X}'} \mu_{\vec{X}} + c_1)(2\sigma_{\vec{X}'\vec{X}}+c_2)}{(\mu_{\vec{X}'}^2 + \mu_{\vec{X}}^2+c_1)(\sigma_{\vec{X}'}^2+\sigma_{\vec{X}}^2 + c_2)},
\end{equation}
where $\mu_{\vec{X}'}$, $\mu_{\vec{X}}$, $\sigma_{\vec{X}'}$ and $\sigma_{\vec{X}}$, are the means and variances of $\vec{X}'$ and $\vec{X}$ respectively, and $\sigma_{\vec{X}'\vec{X}}$ is the covariance between the two. The variables stabilizing the division, $c_1$ and $c_2$ are specified in the same way as \citet{Wang2004}.

\section{Experiments and Results} \label{sec:experiments}
To demonstrate the effectiveness of different configurations, we conducted experiments on images from multiple different datasets, across the four combinations of initialization schemes and distance measures discussed in Section \ref{sec:theory}. We explored the Transformed Gaussian (TG) and uniform (Unif) initialization schemes for the dummy data and the Euclidean (Eucl) and Adaptive Gaussian (AG) distance measures for measuring the difference between gradients.  Experiments were done for two network architectures; LeNet-5 \cite{lenet5} with L-BFGS optimizer \cite{Liu1989} as proposed by \citet{Zhu2020} and ResNet-18 \cite{resnet} with AdamW optimizer \cite{AdamW} as proposed by \citet{Wang2020}. Images from the NIH Chest X-ray dataset provided by NIH Clinical Center were used for the experiments in this section \cite{chestxray8}\footnote{NIH Chest X-ray data: \href{https://nihcc.app.box.com/v/ChestXray-NIHCC}{nihcc.app.box.com/v/ChestXray-NIHCC}}. Further experiments were carried out for the datasets CIFAR-100 \cite{cifar100}, MNIST \cite{mnist}, Omniglot \cite{Lake2015} and SVHN \cite{Netzer2011} (see Appendix \ref{app:recon_other_data}).
\begin{figure}[!t]
    \centering
    \includegraphics[width=\linewidth]{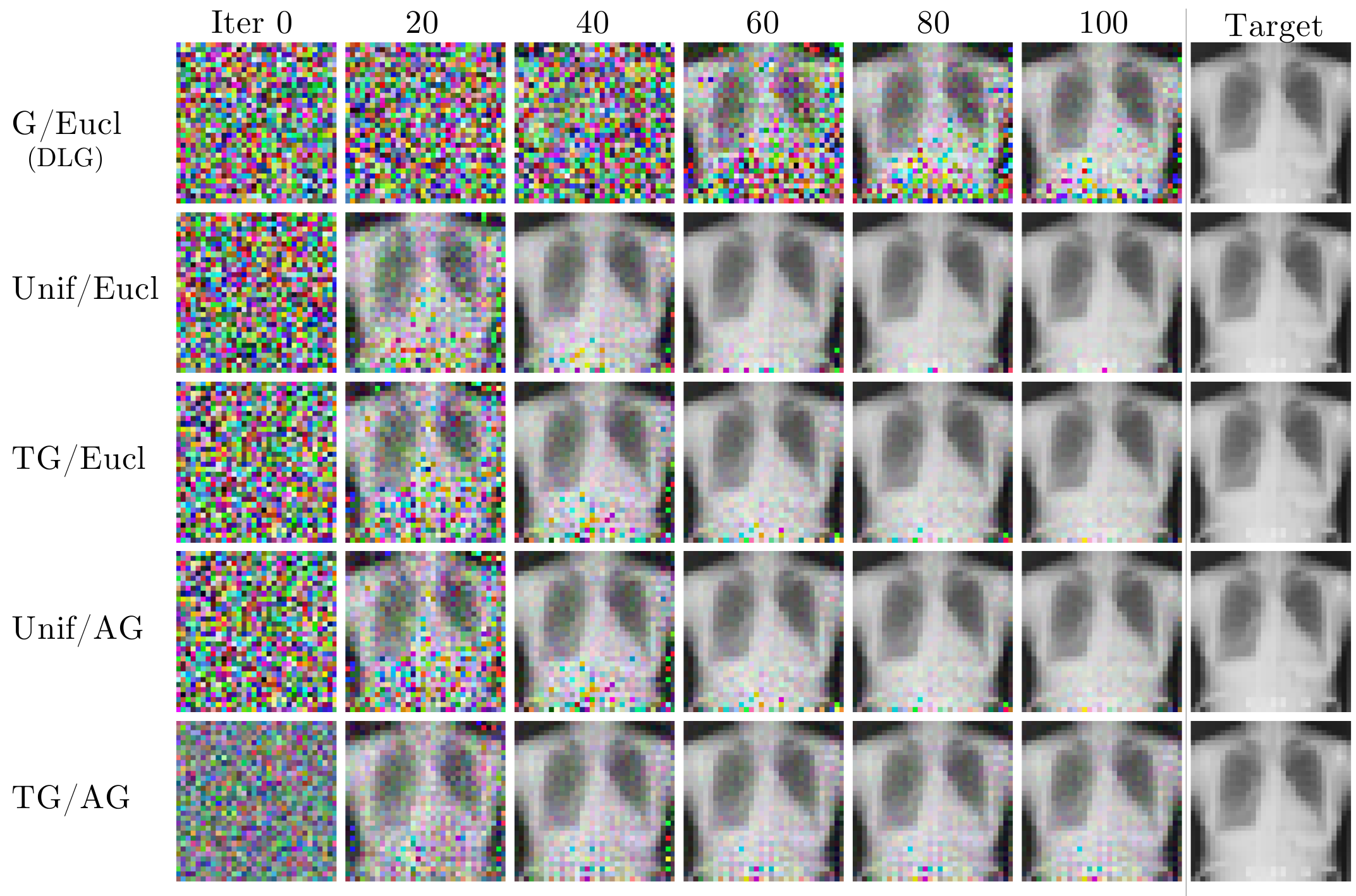}
    \caption{Example of a reconstructed image using L-BFGS optimizer with a learning rate of 0.1 on the LeNet-5 architecture for the different configurations.}
    \label{fig:chest-xray-recon}
\end{figure}

\subsection{Choosing hyper-parameters in the Gaussian measure}
\label{sec:sigma}
\citet{Wang2020} chose $\lambda^2= \Var{\nabla \textbf{W}}$ (Equation \ref{eq:sapag}). However, we found the algorithm to be unstable and rarely produce any identifiable reconstructions on single images using the LeNet-5 architecture. This is likely because $\Var{\nabla \textbf{W}}$ is often very small, which can result in values that are smaller than what can be represented in double precision when the exponential function is applied. We conducted a study of how different values of $\lambda^2$ performed on three metrics; MSE, SSIM, and ability to converge. The results show that choosing a constant value for $\lambda^2$ introduces a trade-off between the three metrics (see Table \ref{tab:sigma-xray-comparison}). For the X-ray dataset, we found that using $\lambda^2=200$ generally performs best in terms of MSE and SSIM. However, using $\lambda^2 = n\text{Var}(\nabla \textbf{W})$ resulted in a better trade-off than for a fixed $\lambda^2$. This is emphasized when reconstructing images from other datasets (see Appendix \ref{app:sigma_choice}). Consequently, specific values of $\lambda^2$ can outperform the AG measure for specific datasets, but the AG measure provides a nice baseline across datasets and at the same time avoids an additional free parameter. The value of $Q$ in equation \ref{eq:sapag} is specified as $1/i$ where $i$ is the layer number such that more weight is applied to layers closer to the input similar to \citet{Wang2020}.
\begin{table*}
\centering
\caption{Average MSE, SSIM and number of non-converging (NNC) of 100 images in the ChestXray dataset using the Gaussian measure with different values of $\lambda^2$ for the two initialization schemes after 100 iterations.}
\begin{tabular}{@{}llrrrrrrrrr@{}}
    \toprule
    & & \multicolumn{9}{c}{$\lambda^2$} \\ \cmidrule(lr){2-10}
    & & \multicolumn{1}{c}{$50$} & \multicolumn{1}{c}{$100$} & \multicolumn{1}{c}{$150$} & \multicolumn{1}{c}{$200$}& \multicolumn{1}{c}{$500$} & \multicolumn{1}{c}{$800$} & \multicolumn{1}{c}{$1000$} & \multicolumn{1}{c}{$1500$} & AG \\ \midrule
    MSE & TG & 8.6e-02 & 3.8e-02 & 5.1e-03 & \textbf{4.7e-03} & 6.1e-03 & 8.6e-03 & 7.2e-03 & 1.1e-02 & 9.6e-03 \\
    & Unif & 1.4e-01 & 5.9e-02 & 1.8e-02 & 1.4e-02 & \textbf{8.3e-03} & 9.3e-03 & 1.7e-02 & 1.1e-02 & 2.1e-02 \\\cmidrule{1-11}
    SSIM & TG & $0.11$ & $0.67$ & $0.87$ & $\bm{0.88}$ & $0.85$ & $0.83$ & $0.83$ & $0.81$ & $0.85$ \\
    & Unif & $0.07$ & $0.54$ & $0.78$ & $\bm{0.83}$ & $0.81$ & $0.79$ & $0.77$ & $0.76$ & $0.81$ \\ \cmidrule{1-11}
    NNC & TG & 36 & 30 & 26 & 19 & 7 & 6 & \textbf{4} & 6 & 10 \\
    & Unif & 34 & 19 & 26 & 24 & 10 & 3 & 8 & \textbf{3} & 8 \\
    \bottomrule
\end{tabular}
\label{tab:sigma-xray-comparison}
\end{table*}

\subsection{Comparing configurations}
An example of image reconstructions from the ChestXray dataset using the four different configurations is shown in Figure  \ref{fig:chest-xray-recon}. However, Figure \ref{fig:chest-xray-recon} only shows a single image and we found that the convergence and quality of reconstructions vary significantly for different images. Examples of more reconstructions are shown in Appendix \ref{app:additional-reconstructions}, where it is also seen that the algorithm sometimes fails to reconstruct images. The numbers of failed reconstructions out of 100 images from the ChestXray dataset are shown in Table \ref{tab:le_non_converge}.
\begin{table}
\centering
\caption{Number of non-converging images using LeNet-5 and L-BFGS. Tests were run for 100 images from the ChestXray dataset. See Appendix \ref{app:recon_other_data} for other datasets.}
\begin{tabular}{@{}lrrcrr@{}}
    \toprule
    & \multicolumn{2}{c}{$lr=1$} & \phantom{M} & \multicolumn{2}{c}{$lr=0.1$} \\ \cmidrule(lr){2-3} \cmidrule(l){5-6}
     & Eucl & AG & & Eucl & AG \\ \midrule
    TG & 21 & 13 & & $5$ & $10$\\
    Unif & 27 & 10 & & $8$ & $8$  \\ \bottomrule
\end{tabular}
\label{tab:le_non_converge}
\end{table}
A similar comparison was made for the ResNet-18 architecture using 20 images, but in this case all configurations successfully reconstruct all 20 images. This indicates that the algorithm is more stable for more complex networks. Table \ref{tab:le_non_converge} also indicates that using a lower learning rate results in a more stable convergence. Choosing the learning rate is in this case a balance between convergence speed and stability. In the following, a learning rate of 0.1 is used for the L-BFGS optimizer and 0.001 for the AdamW optimizer. Experiments are averaged over 100 images for the LeNet-5 architecture and 20 images for the ResNet-18 architecture in order to determine the general performance of different configurations. 
\begin{figure*}[!t]
    \centering
    \includegraphics[width=0.75\linewidth]{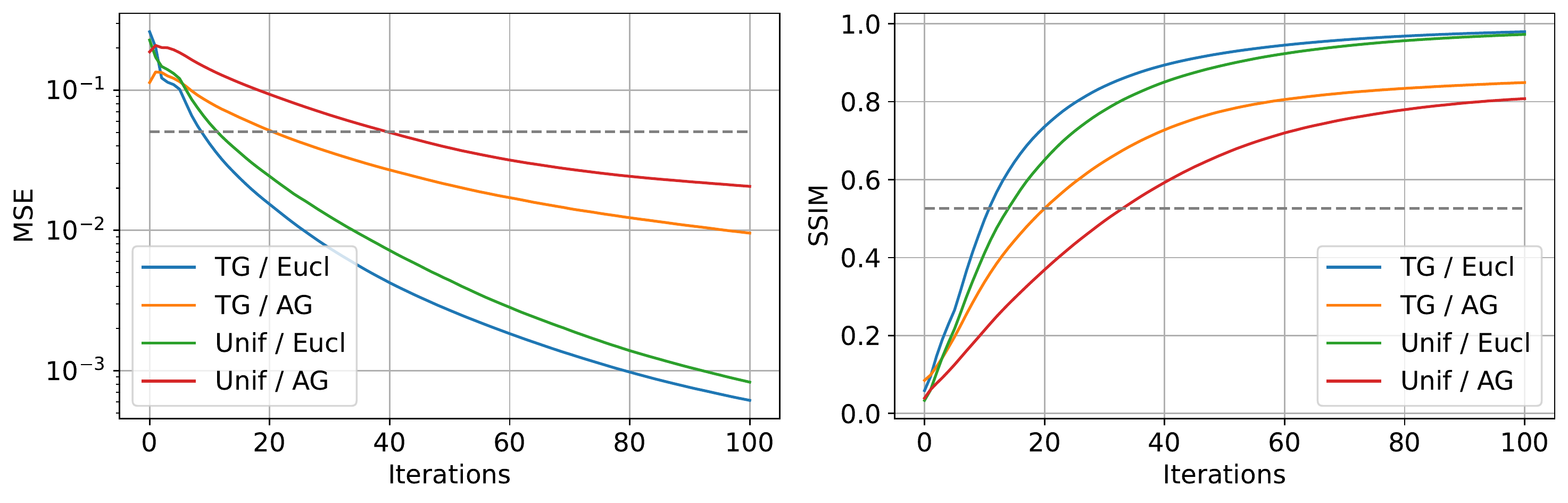}
    \caption{Average reconstruction similarity over 100 images with non-converging images removed (see Table \ref{tab:le_non_converge}) using LeNet-5 with L-BFGS optimizer ($lr=0.1$) on the ChestXray dataset. See Appendix \ref{app:recon_other_data} for reconstruction similarities of other datasets. The grey dotted lines specify baselines for MSE (0.050) and SSIM (0.526) using an algorithm that returns a random image from the dataset.}
    \label{fig:le_convergence_comparison}
\end{figure*}
\begin{figure*}[!t]
    \centering
    \includegraphics[width=0.75\linewidth]{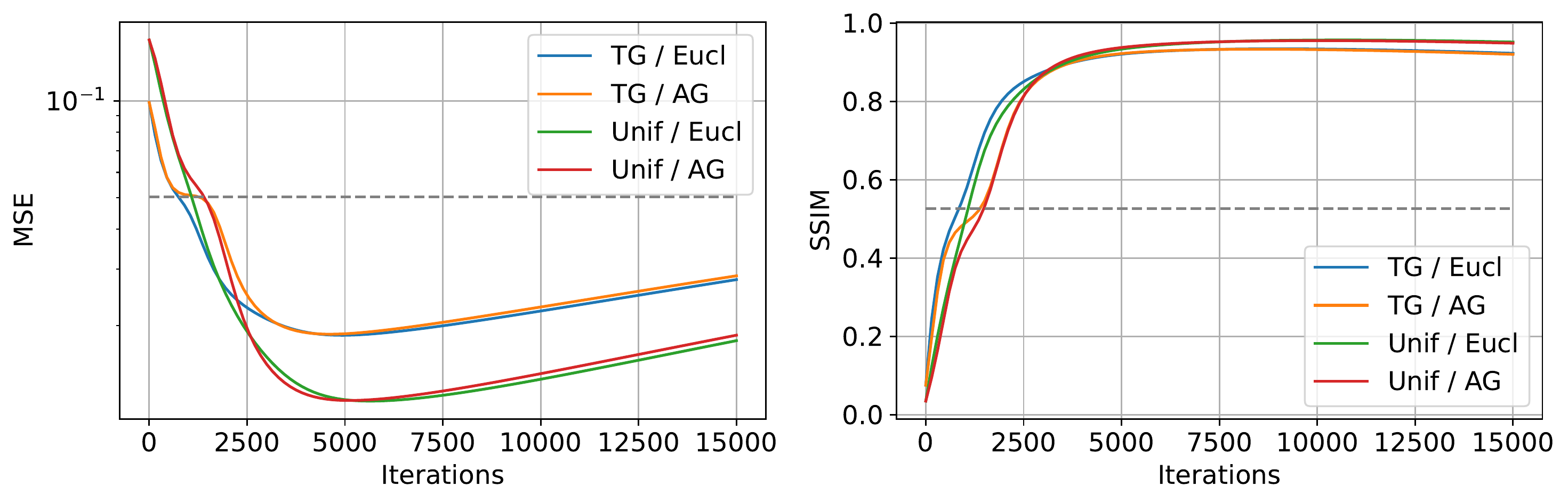}
    \caption{Average reconstruction similarity over 20 converged images using ResNet-18 with AdamW optimizer ($lr = 0.001$) on the ChestXray dataset. The grey dotted lines specify baselines for MSE (0.050) and SSIM (0.526) using an algorithm that returns a random image from the dataset.}
    \label{fig:res_convergence_comparison}
\end{figure*}

SSIM and MSE metrics are not always well suited in the medical domain since medical images are often processed to be similar as opposed to natural images. Therefore, we provide a baseline based on an algorithm that simply returns a random but different image from the dataset. We find that all four configurations beat these baselines.

The results using LeNet-5 on the ChestXray dataset are shown in Fig.\ \ref{fig:le_convergence_comparison}. Generally, the Euclidean distance measure improves both reconstruction speed and quality. The Adaptive Gaussian measure does not converge as fast as the Euclidean measure. Additionally, it becomes evident that the Transformed Gaussian initialization performs better than the uniform initialization when distance measures are the same.

The results using ResNet-18 on the ChestXray dataset are shown in Fig.\ \ref{fig:res_convergence_comparison}. The final convergence accuracy seems to depend more on the initialization scheme for ResNet-18 than LeNet-5. The uniform initialization results in a lower MSE on average after $5000$ iterations for both distance measures even though the Transformed Gaussian configurations start closer to the true images. Consequently, our results indicate that the distance measures perform equally well on ResNet-18 given an initialization scheme.


\section{Discussion}\label{sec:discussion}
The main purpose of the conducted experiments was to determine how different initialization schemes and distances measures affect the ability to reconstruct possibly sensitive data in a federated learning setting. For the LeNet-5 architecture, the different combinations of initialization schemes and distance measures start out equally well and separate over time as seen in Fig.\ \ref{fig:le_convergence_comparison}, while the opposite scenario is observed for the ResNet-18 architecture in Fig.\ \ref{fig:res_convergence_comparison}. This raises the question of how important the network architecture is in determining an attacker's ability to steal private data as results are not consistent across different architectures. This notion is further supported by the improved stability in reconstruction convergence observed for ResNet-18 where all images converged as opposed to LeNet-5. Whether it is the network architectures themselves, the different optimizers and learning rates used for each of them, or a combination hereof should be investigated further. One should note that reconstruction time on larger networks such as ResNet-18 ($\approx$ 14.5M parameters) is much longer than on smaller networks similar to LeNet-5 ($\approx$ 85,000 parameters). However, larger networks is not a viable defense strategy as shown by the algorithm being able to reconstruct images using the ResNet architecture. It is worth noting that it is still possible to reconstruct the images for different network architectures as long as it satisfies the minimal structural requirements for reconstruction, see \cite{Qian2020}. Furthermore, batches of images can also be reconstructed, but how the different configurations relate to batch reconstructions should be investigated further. The present study shows that the reconstruction stability and quality depends on the choice of initialization scheme and distance measure to a large extent. Consequently, an attack can be drastically improved by searching for and choosing the right configuration. 



\section{Conclusion}\label{sec:conclusion}
We investigated how different initialization schemes and distance measures may improve the convergence speed and quality of the DLG algorithm \cite{Zhu2020}. We introduced the Transformed Gaussian initialization scheme for initializing dummy images, which increases the reconstruction speed and quality of images from different datasets compared to using a uniform distribution using the network architecture LeNet-5. Additionally, we found that introducing a Gaussian distance measure worsens performance compared to a simple Euclidean distance measure for LeNet-5. Furthermore, we found that there is little to no advantage gained from using different initialization schemes and distance measures for the more complex network architecture ResNet-18.

\appendices
\section{Choosing the parameter in the Gaussian measure} \label{app:sigma_choice}
\begin{figure}
\centering
\includegraphics[width=\linewidth]{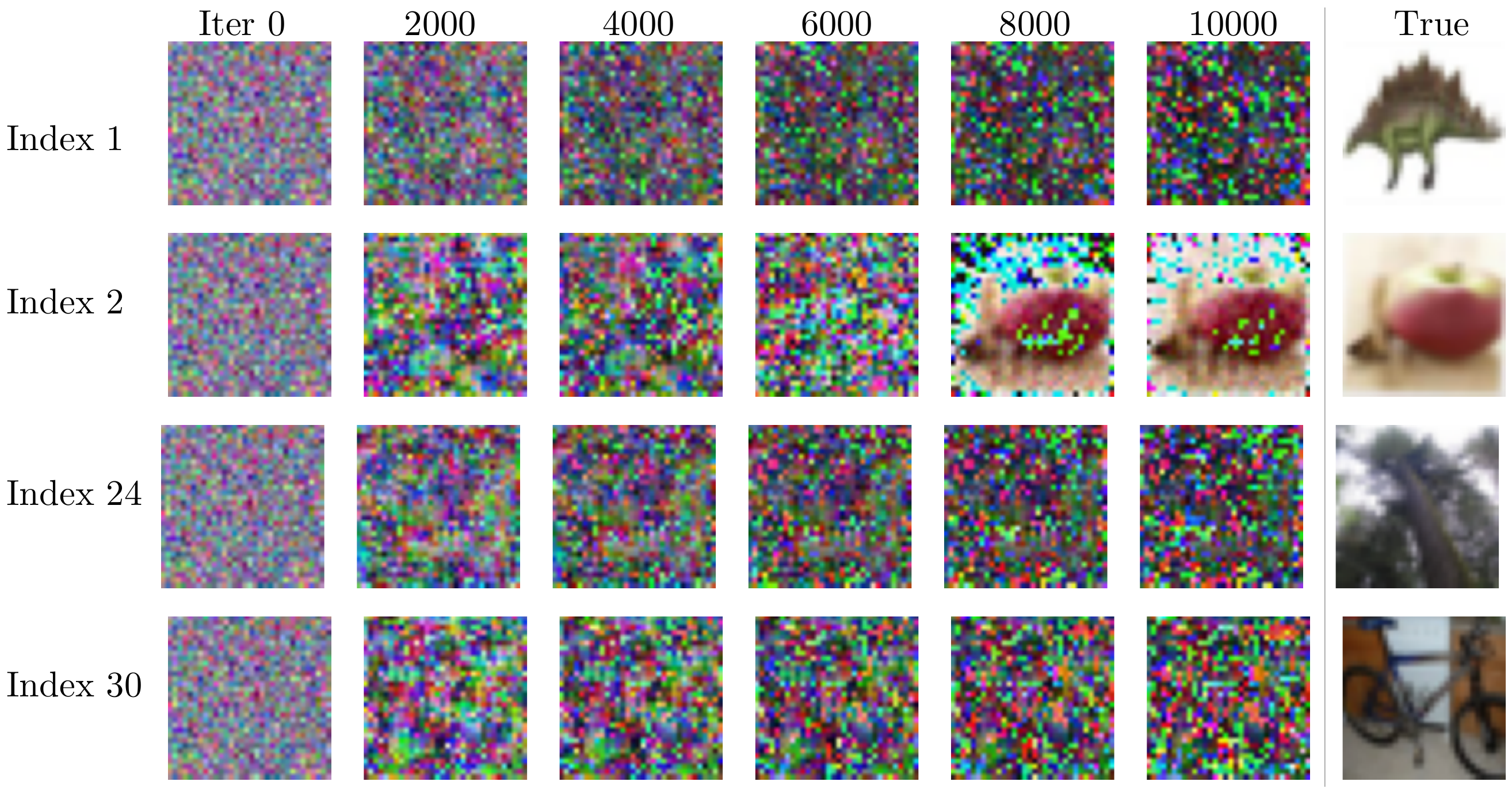}
\caption{Reconstructed images using AdamW optimizer with learning rate of 0.001 on the ResNet-18 architecture using the CIFAR-100 dataset. These images are reconstructed using the Transformed Gaussian initialization scheme and the Gaussian measure from Equation \ref{eq:sapag}. Consequently, it is the same setup as in the SAPAG paper \cite{Wang2020}. We clearly see a slow convergence compared to the reconstruction images from appendix \ref{app:additional-reconstructions}.}
\label{fig:sapag_recon}
\end{figure}

It is necessary to find a feasible value of $\lambda^2$ when working with the Gaussian measure. We conducted experiments using the CIFAR dataset in order to investigate how the algorithm performs for different values of $\lambda^2$. For $\lambda^2> 800$, more images converge but when $150\leq \lambda^2 \leq800$, a higher similarity with the ground truth image is achieved. We found that using a value that is adaptive to the target achieves both stable convergence and high similarity. Assume for some arbitrary layer that $\nabla\textbf{W} \in \mathbb{R}^n$, then we compute $\lambda^2$ in AG as
\begin{equation}
    \lambda^2 = n\text{Var}(\nabla\textbf{W}).
\end{equation}
Using this adaptive measure results in the most accurate reconstructions when using TG initialization on CIFAR-100 (MSE: 5.71e-3, SSIM: 0.850), while still keeping a low number of non-converging images (6). When using the uniform initialization it is slightly worse than for some fixed values of $\lambda^2$ considering MSE and SSIM, but the number of non-converging images is still kept relatively low (7). Consequently, $\lambda^2 = n\text{Var}(\nabla\textbf{W})$ is a reasonable estimate of the optimal $\lambda^2$ and strikes a balance in the trade-off between stability and reconstruction quality. The results are shown in Table \ref{tab:sigma-cifar-comparison}. Furthermore, we found using $\lambda^2=\Var{\nabla \vec{W}}$ as \citet{Wang2020} to be unstable (see Fig. \ref{fig:sapag_recon})
\begin{table*}
\centering
\caption{Average MSE, SSIM and number of non-converging (NNC) of 100 images in the CIFAR dataset using the Gaussian measure with different values of $\lambda^2$ for the two initialization schemes after 100 iterations.}
\label{tab:sigma-cifar-comparison}
\begin{tabular}{@{}llrrrrrrrrrrr@{}}
    \toprule
        & & \multicolumn{11}{c}{$\lambda^2$} \\ \cmidrule(lr){3-12}
        & & \multicolumn{1}{c}{$1$} & \multicolumn{1}{c}{$50$} & \multicolumn{1}{c}{$100$} & \multicolumn{1}{c}{$150$} & \multicolumn{1}{c}{$200$}& \multicolumn{1}{c}{$500$} & \multicolumn{1}{c}{$800$} & \multicolumn{1}{c}{$1000$} & \multicolumn{1}{c}{$1500$} & \multicolumn{1}{c}{$2000$} & AG \\ \midrule
        MSE & TG & N/A & 3.24e-1 & 1.37e-1 & 4.47e-2 & 1.37e-2 & 1.99e-2 & 6.44e-3 & 1.52e-2 & 7.60e-3 & 8.05e-3 & \textbf{5.71e-3} \\
        & Unif & N/A & 1.54e-1 & 7.36e-2 & 1.28e-2 & \textbf{6.48e-3} & 7.80e-3 & 1.35e-2 & 1.05e-2 & 1.20e-2 & 1.29e-2 & 9.96e-3  \\\cmidrule{1-13}
        SSIM & TG & N/A & $0.012$ & $0.483$ & $0.726$ & $0.783$ & $0.846$ & $0.833$ & $0.707$ & $0.814$ & $0.802$ & $\textbf{0.850}$ \\
        & Unif & N/A & $0.054$ & $0.553$ & $0.831$ & $\textbf{0.851}$ & $0.809$ & $0.776$ & $0.778$ & $0.750$ &  $0.734$ & $0.783$ \\\cmidrule{1-13}
        NNC & TG & 100 & 35 & 15 & 18 & 14 & 12 & 6 & \textbf{3} & \textbf{3} & 5 & 6 \\
        & Unif & 100 & 24 & 16 & 27 & 14 & 7 & 5 & 8 & \textbf{3} & 10 & 7 \\
        \bottomrule
    \end{tabular}
\end{table*}

\section{Reconstructions of different data sets} \label{app:recon_other_data}
\begin{figure*}
\subfigure[CIFAR-100]{
    \label{fig:cifar}
    \includegraphics[width=0.45\linewidth]{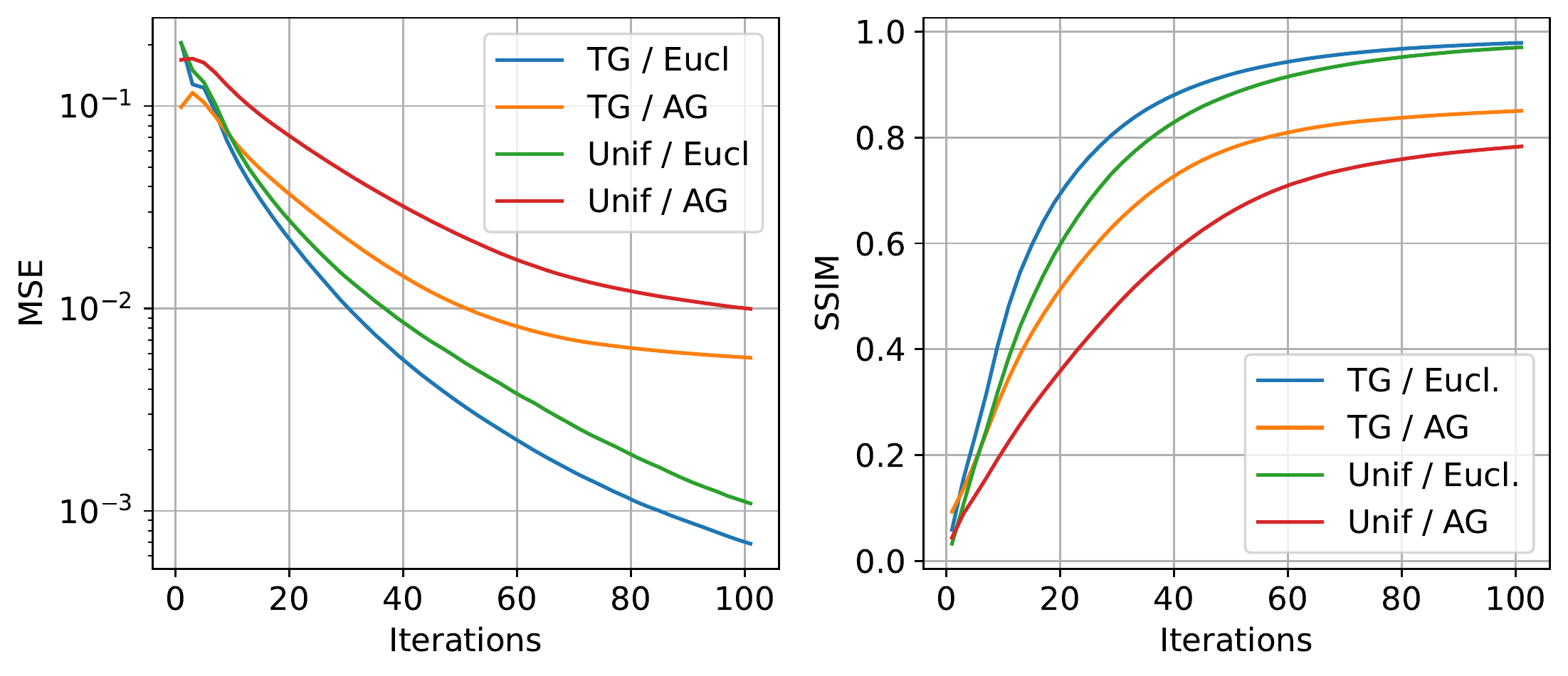}
}\hfill
\subfigure[MNIST]{
    \label{fig:mnist}
    \includegraphics[width=0.45\linewidth]{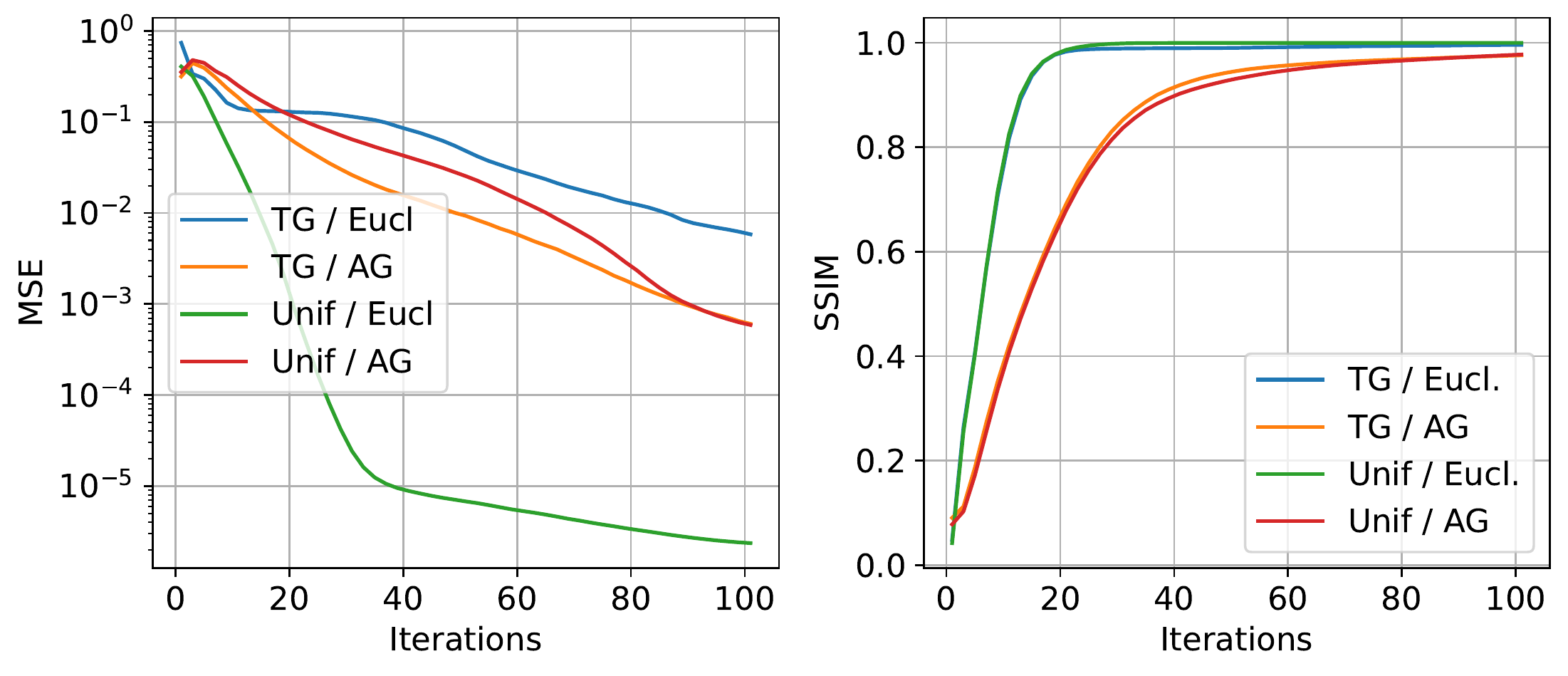}
} \hfill
\subfigure[Omniglot]{
    \label{fig:omniglot}
    \includegraphics[width=0.45\linewidth]{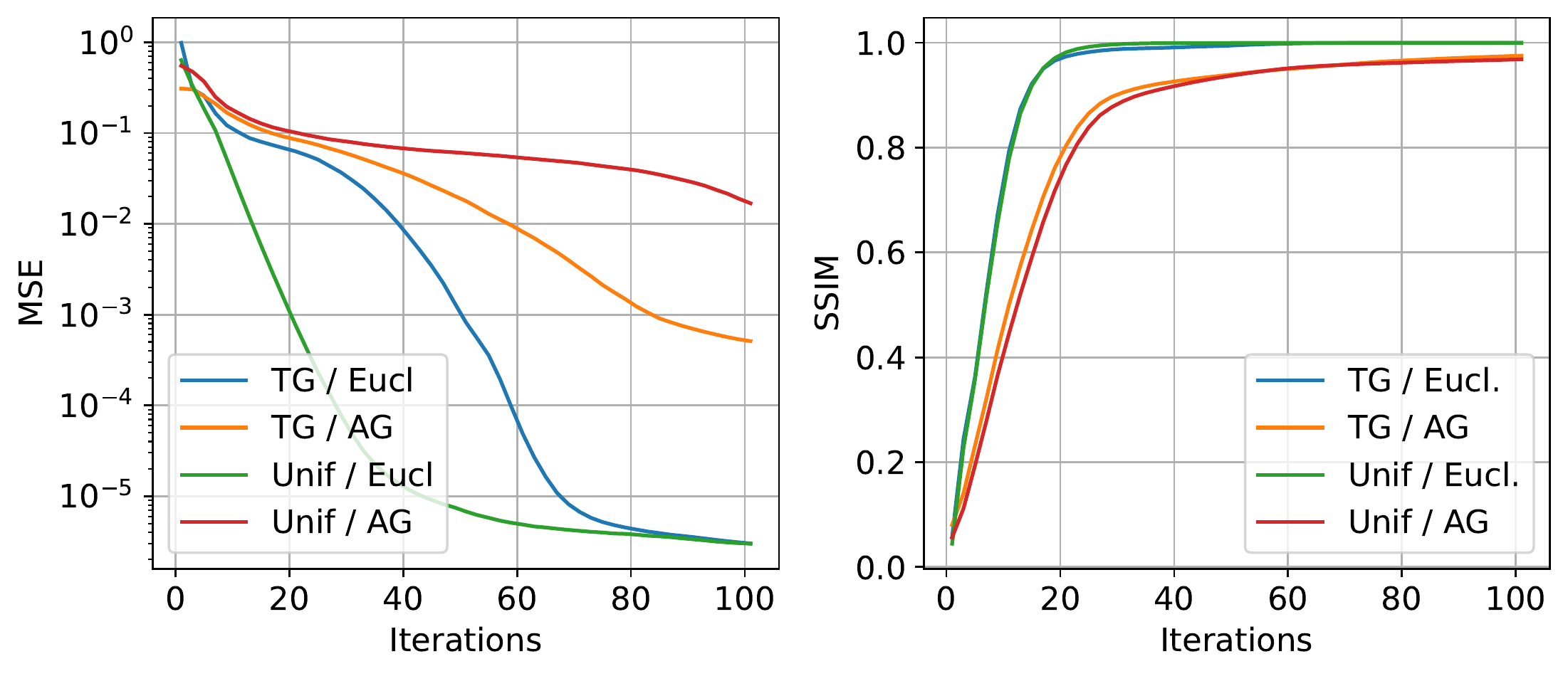}
}\hfill
\subfigure[SVHN]{
    \label{fig:svhn}
    \includegraphics[width=0.45\linewidth]{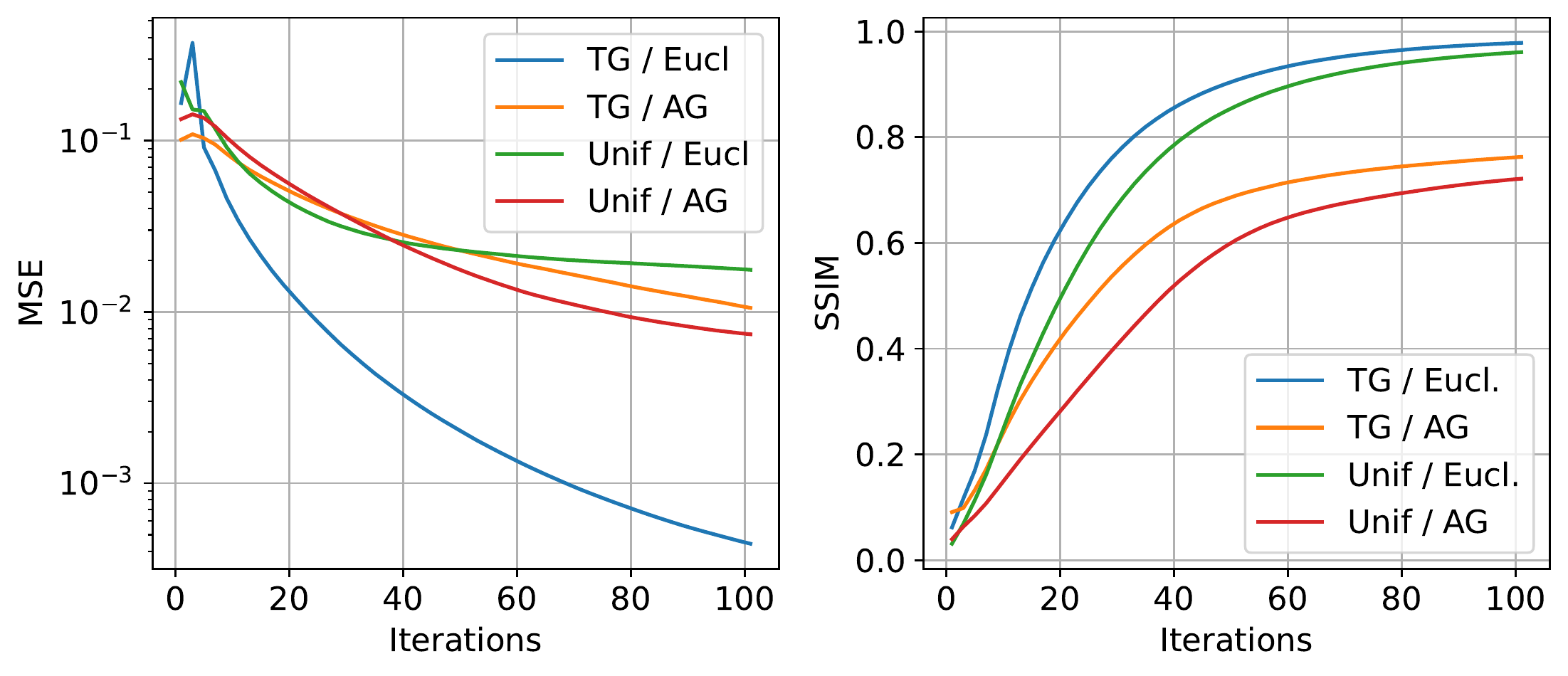}
}
\caption{Average reconstruction similarity for the converged images out of 100 images of different data sets.}
\label{fig:convergence_other_data}
\end{figure*}

We conducted similar experiments for the datasets CIFAR-100 \cite{cifar100}, MNIST \cite{mnist}, Omniglot \cite{Lake2015}, and SVHN \cite{Netzer2011}. Convergence plots are shown in Fig.\ \ref{fig:convergence_other_data} and the number of non-converging images are shown in Table \ref{tab:le_non_converge_other_data}. For MNIST and Omniglot data, we expected the data to be more uniformly distributed than Gaussian since the pixels are mostly black or white and it is also the configuration Unif/Eucl that performs best for those datasets. For SVHN and CIFAR data, the case is the opposite, where we expect data to be closer to a Gaussian distribution. Note that SSIM indicates that using euclidean distance measure is faster than the adaptive Gaussian measure. In all cases, the initialization scheme does not seem to have as big an impact on the final convergence when using the adaptive Gaussian measure.
\begin{table}
\centering
\caption{Number of non-converging images for all combinations of initialization schemes and distance measures in the LeNet-5 architecture using L-BFGS with $lr=0.1$. Tests were run for 100 images from each data set.}
\begin{tabular}{lrrrrrrrr}
    \toprule
    & \multicolumn{2}{c}{CIFAR} & \multicolumn{2}{c}{MNIST} & \multicolumn{2}{c}{Omniglot} & \multicolumn{2}{c}{SVHN}\\
    \cmidrule(lr){2-3} \cmidrule(lr){4-5} \cmidrule(lr){6-7} \cmidrule(lr){8-9}
            & Eucl & AG & Eucl & AG & Eucl & AG & Eucl & AG \\ \midrule
    TG      & 4 & 2 & 4 & 8 & 3 & 7 & 4 & 4 \\
    Unif    & 5 & 13 & 4 & 8 & 7 & 10 & 4 & 3 \\ \bottomrule
\end{tabular}
\label{tab:le_non_converge_other_data}
\end{table}

\section{Additional reconstructed images}\label{app:additional-reconstructions}
\begin{figure*}
\subfigure[CIFAR]{
    \label{fig:CIFAR3}
    \includegraphics[width=0.45\linewidth]{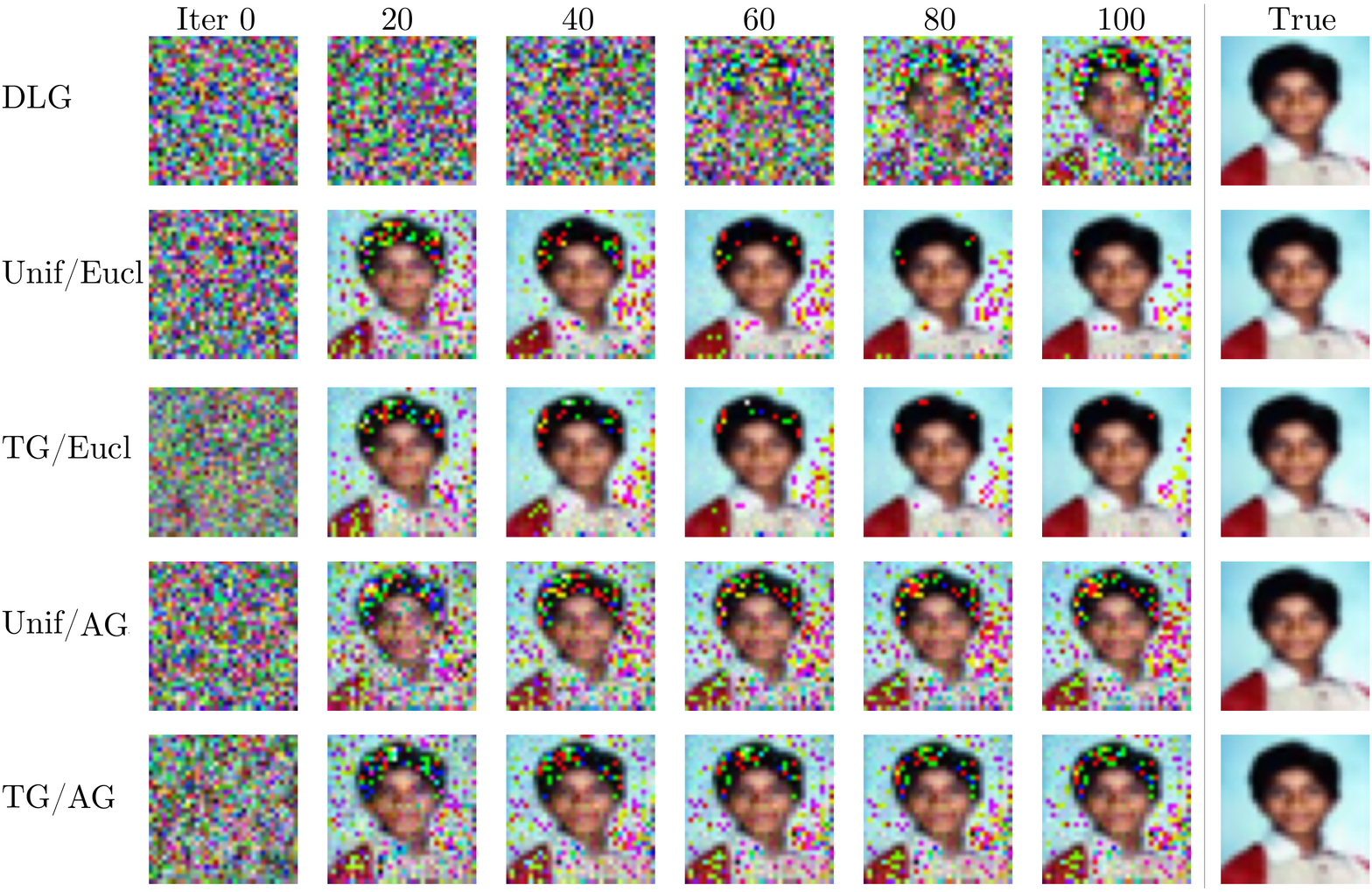}
}\hfill
\subfigure[MNIST]{
    \label{fig:mnist_30}
    \includegraphics[width=0.47\linewidth]{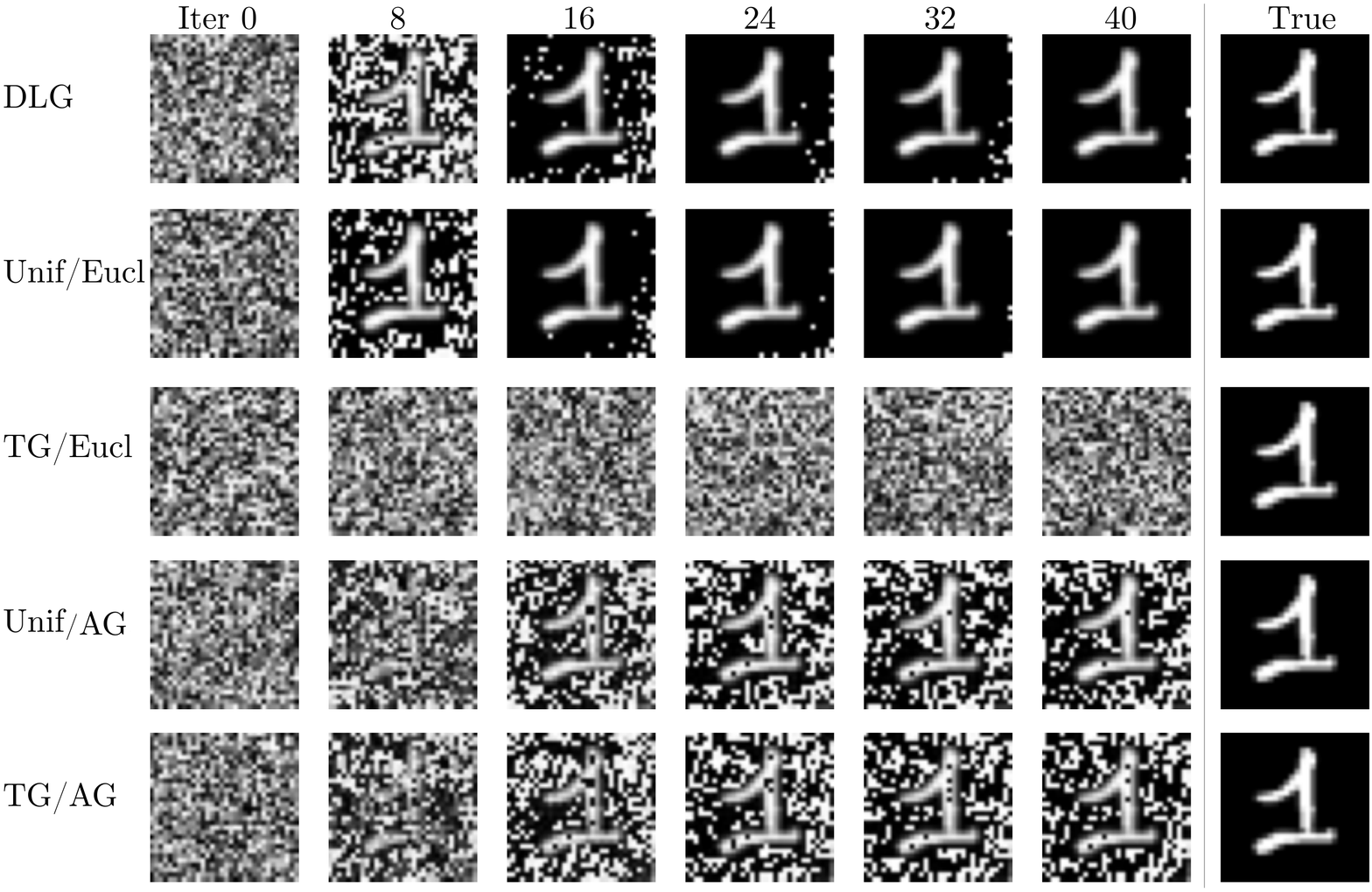}
}
\subfigure[Omniglot]{
    \label{fig:Omni10}
    \includegraphics[width=0.47\linewidth]{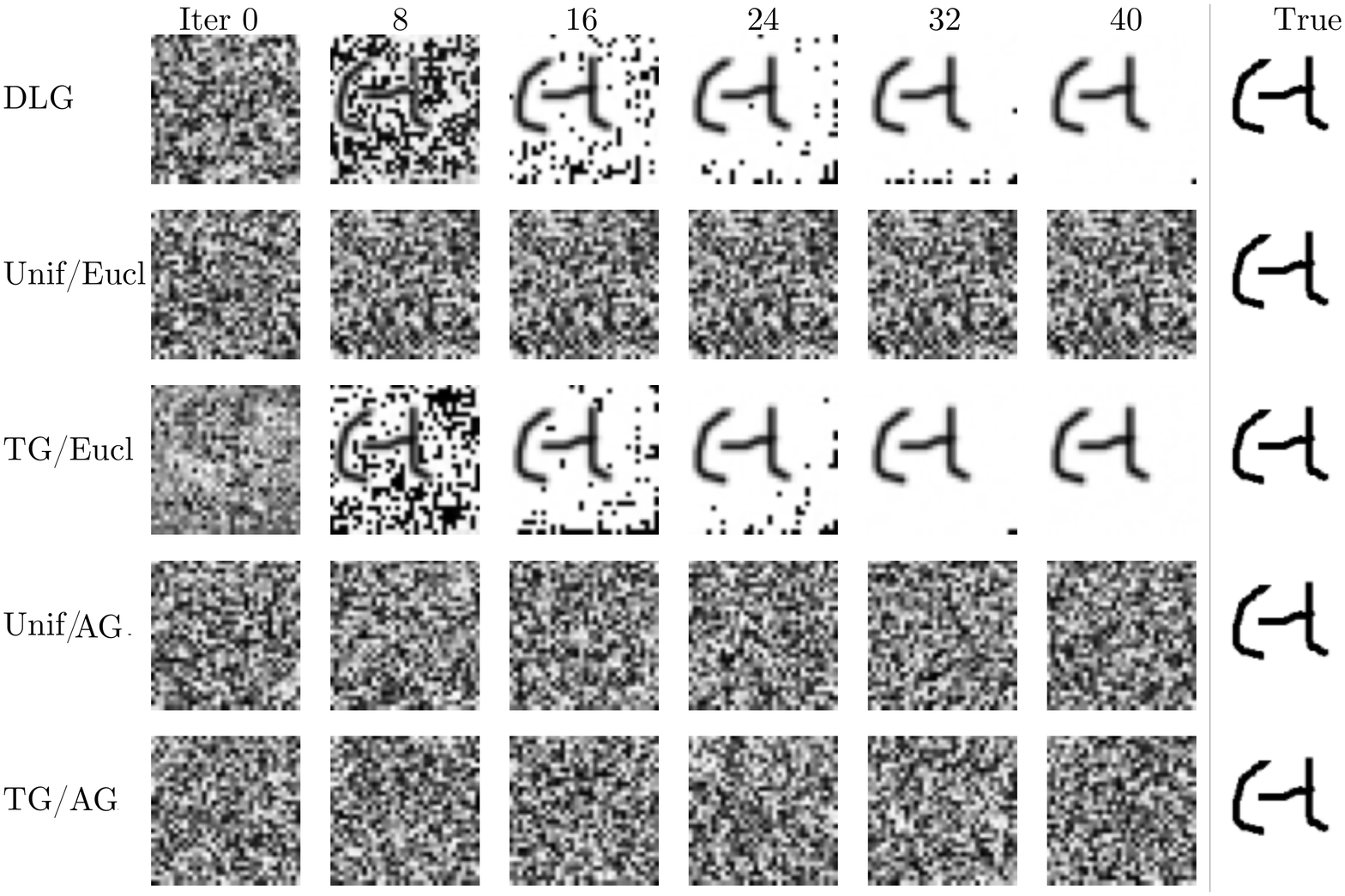}
} \hfill
\subfigure[SVHN]{
    \label{fig:SVHN110}
    \includegraphics[width=0.47\linewidth]{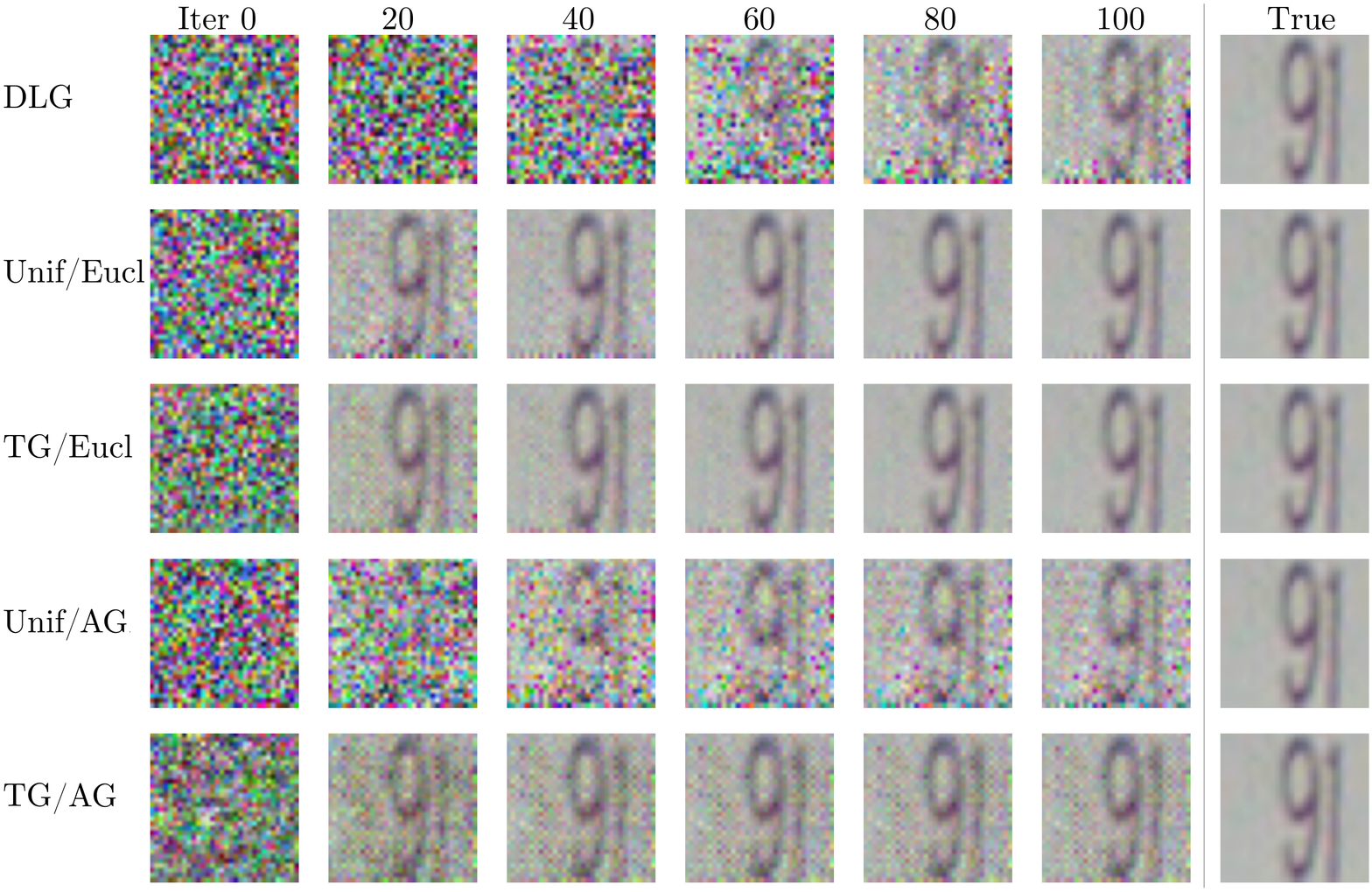}
}
\caption{Reconstructed images from different datasets dataset using L-BFGS optimizer with learning rate of 1 on the LeNet-5 architecture. Different configurations seem to be struggling on different images.}
\label{fig:other-reconstructions}
\end{figure*}

\begin{figure*}
\subfigure[]{
    \label{CIFAR24_resnet}    
    \includegraphics[width=0.48\linewidth]{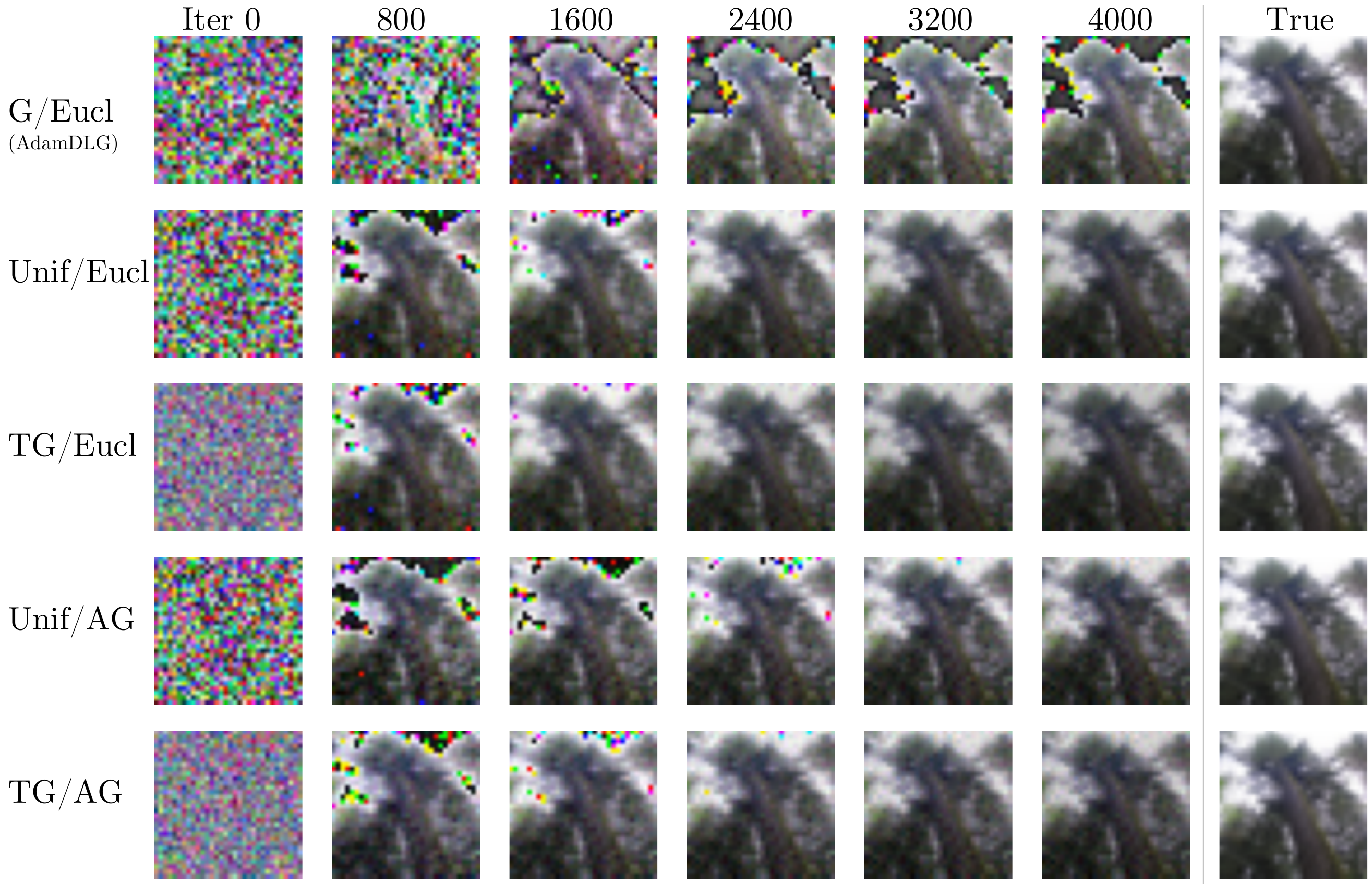}
}\hfill
\subfigure[]{
    \label{CIFAR30_resnet}
    \includegraphics[width=0.48\linewidth]{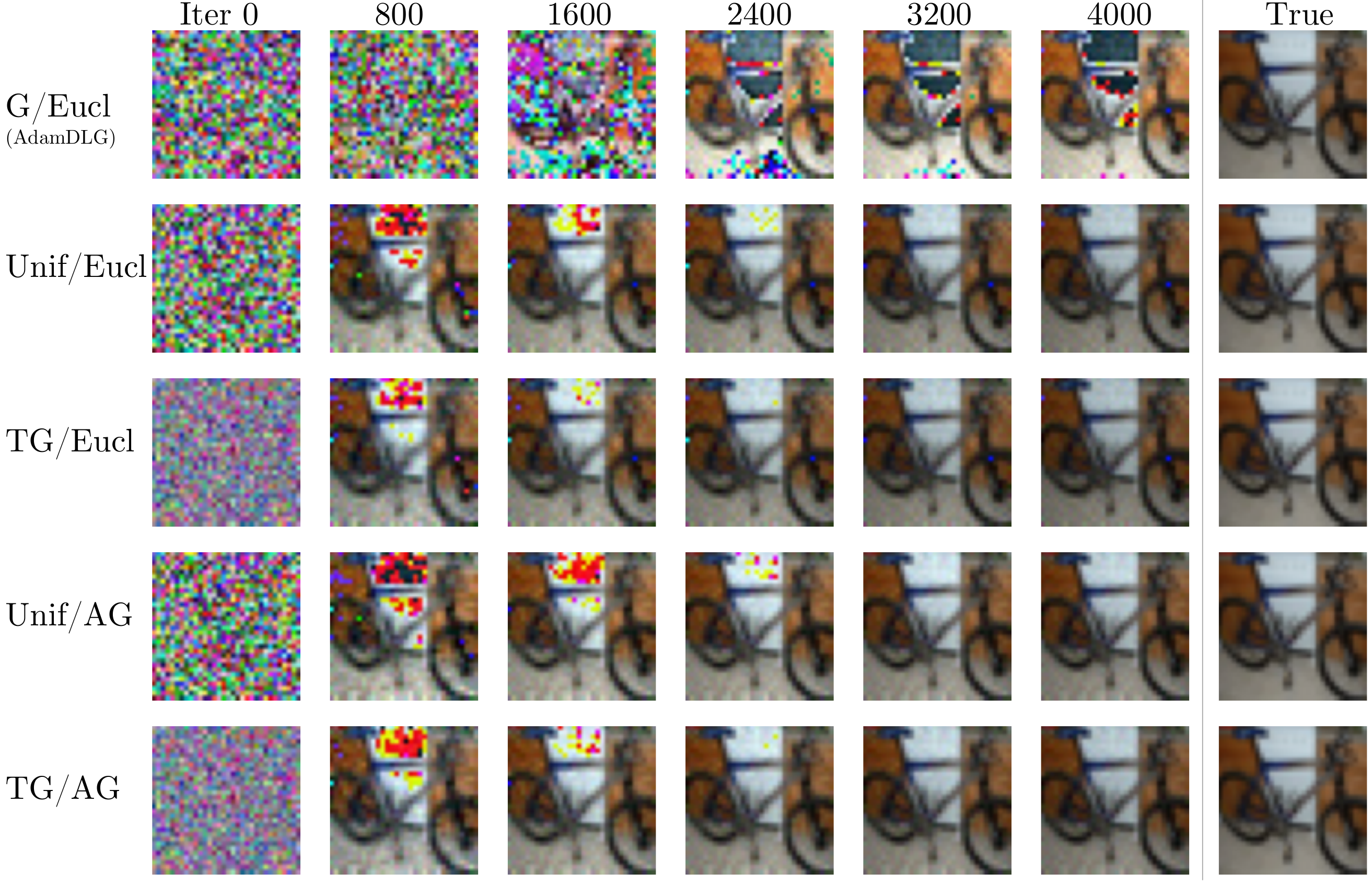}
}
\caption{Reconstructed images from the CIFAR-100 dataset using AdamW optimizer with learning rate of 0.001 on the ResNet-18 architecture.}
\label{fig:resnet18-cifar-reconstructions}
\end{figure*}

Examples of reconstructed images from multiple different datasets are shown in Fig. \ref{fig:other-reconstructions}. Furthermore, reconstructions images using the ResNet architecture are shown in Fig. \ref{fig:resnet18-cifar-reconstructions}. In these examples the original DLG configuration is outperformed by all of the four proposed configurations for most images. Interestingly, some configurations are able to converge for some images whereas other configurations seem to perform better on other images. This further indicates that the choice of configuration including optimizer and dataset has a large impact on the quality and stability of reconstructions. However, it is still unclear if there is a correlation between the image data distribution and optimal choice of configuration. This raises the question, for further research, of whether specific configurations perform better on specific types of images.

\newpage
\newpage
\bibliographystyle{IEEEtranN}
\bibliography{bibliography}{}

\end{document}